\documentclass[dvips]{acta}
\usepackage{supertabular,lscape,epsfig}
\usepackage{amssymb}
\usepackage{amsmath}

\SetPages{0}{0}

\SetVol{66}{2016}

\begin{document}

\begin{Titlepage}
\Title{The OGLE Collection of Variable Stars.\\Eclipsing Binaries in the Magellanic System\footnote{Based on observations
obtained with the 1.3-m Warsaw telescope at the Las Campanas Observatory of
the Carnegie Institution for Science.}}
\Author{M.~~P~a~w~l~a~k$^1$,~~
I.~~S~o~s~z~y~{\'n}~s~k~i$^1$,~~
A.~~U~d~a~l~s~k~i$^1$,~~
M.\,K.~~S~z~y~m~a~{\'n}~s~k~i$^1$,~~\\
\L.~~W~y~r~z~y~k~o~w~s~k~i$^1$,~~
K.~~U~l~a~c~z~y~k$^2$,~~
R.~~P~o~l~e~s~k~i$^{1,3}$,~~\\
P.~~P~i~e~t~r~u~k~o~w~i~c~z$^1$,~~
S.~~K~o~z~\l~o~w~s~k~i$^1$,~~
D.~M.~~S~k~o~w~r~o~n$^1$,~~\\
J.~~S~k~o~w~r~o~n$^1$,~~
P.~~M~r~{\'o}~z$^1$,~~and~~
A.~~H~a~m~a~n~o~w~i~c~z$^1$
}
{$^1$Warsaw University Observatory, Al.~Ujazdowskie~4, 00-478~Warszawa, Poland\\
e-mail: (mpawlak,soszynsk,udalski)@astrouw.edu.pl\\
$^2$Department of Physics, University of Warwick, Gibbet Hill Road, \\Coventry, CV4 7AL, UK \\
$^3$Department of Astronomy, Ohio State University, 140 W. 18th Ave., Columbus, OH~43210, USA}

\Received{Month Day, Year}
\end{Titlepage}

\Abstract{We present the collection of eclipsing binaries in the Large and Small Magellanic Clouds, based
on the OGLE survey. It contains $48\;605$ systems, $40\;204$ belonging to the LMC and $8\;401$ to the SMC. Out of the total
number of the presented binaries, $16\;374$ are the new ones. We present the time-series photometry obtained for the 
selected objects during the fourth phase of the OGLE project.
The catalog has been created using a two step machine learning procedure based on the Random Forest algorithm.}
{Stars: binaries: eclipsing, Stars: variables:general, Magellanic Clouds}

\section{Introduction}

Binary stars are interesting in astrophysical studies for various reasons. They are an extremely useful
tool for the determination of physical parameters of individual stars (\eg Pietrzy{\'n}ski \etal 2010, He{\l}miniak \etal 2015, Gallena \etal 2016).
Moreover, eclipsing binaries can also serve as a precise distance ruler (\eg Pie-trzy{\'n}ski \etal 2013, Graczyk \etal 2014). 

The Magellanic System, composed of the Large (LMC) and Small Magellanic Clouds (SMC) -- two satellite galaxies of the Milky Way -- is a perfect 
environment for the stellar variability studies. Eclipsing binaries in this region have already been analyzed in the previous 
phases of the Optical Gravitational Lensing Experiment (OGLE). The catalogs from the second phase of the project (OGLE-II) have been prepared by Udalski \etal (1998) and Wyrzykowski \etal (2004) 
in the SMC and by Wyrzykowski \etal (2003) in the LMC. Graczyk \etal (2011) and Pawlak \etal (2013) presented
the OGLE-III catalogs of eclipsing binaries in the LMC and SMC, respectively. Search for varible stars in the Gaia South Ecliptic Pole,
located in the outskirts of the LMC resulted in discovery of 1377 eclipsing binaries (Soszy{\'n}sk \etal 2012).
Large samples of this type of variable stars in the Magellanic System have also been published by Alcock \etal (1997), Faccioli \etal (2005)  
and Derekas \etal (2007) based on the MACHO project, as well as by Muraveva \etal (2014) and Kim \etal (2014) from the EROS project data.

The identification of a statistically significant samples of binary systems in the LMC enabled the calibration of 
the period--luminosity relations formed by contact or close systems. Relations for ellipsoidal red giants have been studied
by Wood \etal (1999), Rucinski and Maceroni (2001), Soszy{\'n}ski \etal (2004) and Pawlak \etal (2014). Period-luminosity-color
relation formed by early-type, massive contact binaries has been recently presented by Pawlak (2016).

In this work, we present the OGLE collection of binary stars in the Magellanic System, containing $48\;605$ objects including
$40\;204$ systems in the LMC, and $8\;401$ in the SMC. The structure of the paper is as follow. Section~2 describes 
the technical details of the instrumental setup, observations and data reduction. Section~3 presents the 
machine learning procedure based on the Random Forest algorithm used to select the objects. The collection of the eclipsing binaries in the Magellanic System 
is presented in Section~4. Section~5 contains the discussion of the results. Section~6 describes a system with a Type II Cepheid as one of the components, 
and Section~7 summarizes the results. 

\section{Observations and Data Reduction}

All the data used in this work were collected during the fourth phase of the OGLE project, between March 2010 and July 2015,
using the 1.3-m Warsaw telescope located at Las Campanas Observatory (operated by the Carnegie Institution for Science) in Chile. The telescope is
equipped with a 32-chip mosaic camera, with a total field of view of 1.4 square degrees. The whole area of 
the Magellanic Clouds region covered by 475 OGLE-IV fields is about 650 square degrees.

Observations were carried out in the {\it I}- and {\it V}-band filters, with about 90\% taken in the {\it I}-band.
The {\it I}-band magnitude range covered by OGLE-IV is from 13 to 21.5~mag. 
The number of collected epochs varies between 100 and 700. The photometry
was obtained with the Differential Image Analysis method (Alard and Lupton 1998, Wozniak 2000). 
For technical details of the OGLE-IV instrumental setup and date reduction system we refer the reader to Udalski \etal (2015a).
The procedure of the photometric uncertainties correction is presented in Skowron \etal (2016)

\section{Machine Learning Classification}

\subsection{Period determination}

The first problem in the process of binary stars selection is the period determination.
The Fourier-transform-based algorithms, like Lomb-Scarlge periodogram (Scarlge 1982)
commonly used for this purpose (\eg Kim \etal 2014), work well for pulsating
variables. However, they often fail in the case of strongly non-sinusoidal light curves like
those of eclipsing stars. Detached binaries with very narrow eclipses are especially
problematic in this case.

Therefore, the Box-fitting Least Squares (BLS) algorithm (Kov{\'a}cs \etal 2002) was used to determine the period for each {\it I}-band light curve collected by
OGLE in the area of the Magellanic Clouds. This method, which was designed to look for transit variability,
seems to be better suited for the search for eclipsing binaries. Another benefit
of using BLS is that apart from the period, the algorithm also provides a series
of parameters of the light curve which can be used as a features in the machine learning
classification process.

For all of the objects used in this study the periods are determined with the BLS algorithm, with the
search range between 0.04~d and 1000~d, with 50000 steps in the frequency range. The minimum and
maximum value of the fraction of points in eclipse $q$ are set to 0.01 and 0.3, respectively.
We used the BLS implementation from VARTOOLS Light Curve Analysis Program (Hartman and Bakos 2016).

\subsection{First Step}

With the periods determined for all of the objects, the proper machine learning 
classification was a next step. For that purpose the Random Forest (Breiman 2001) was used.
To perform this type of classification, a training set needs to be constructed first.
The OGLE-III catalog of eclipsing binaries in the LMC (Graczyk \etal 2011)
was used for this purpose. The entire sample of known eclipsing binaries
from one of the OGLE-IV fields (LMC501) containing 1017 object was
complemented with a matching number (1024) of other, randomly selected stars from the
same field.
The BLS statistics (Tab.~1) were used as features for Random Forest.
The implementation of this algorithm from the WEKA data mining software package (Hall \etal 2009)
was used.

The overall performance of the classifier was evaluated with a 10-fold cross-validation method,
resulting in the average accuracy of 97.6\%.
To describe the performance of the method more precisely, two parameters
were used. To measure the completeness of the search, we defined a {\it
recall} parameter, which is the ratio of the number of true positives to
the entire number of all binaries. The purity of the sample is measured
by the {\it precision} parameter, defined as the ratio of the true
positives to all positive detections.
For the first step of the classification both of these parameters were high: 97\% and 98\%, respectively.
While the {\it recall} level can be considered as satisfactory, the {\it precision} of the search had to be improved, because
with millions of objects to be classified, even a 2\% false positive ratio
can generates a very large number of false detections. Therefore, further steps were implemented to
perform classification more effectively.

To reduce the number of false detections, the cut on the probability is introduced. The positive
detection threshold is increased from default 50\% to 80\%. This step decreases slightly the
{\it recall} of the classification to 93\%, but it allows us to eliminate the large number of false detections
which is essential when classifying the whole spectrum of observed objects, out of which most are
not eclipsing binary stars.

\subsection{Second Step}

At the beginning of the second step of the classification, the training set is composed of the objects,
that passed the first step -- both true and false detections. The ratio of true to false
positives in the output of the first step is about 1:6.
Apart from the BLS parameters, a series of statistical features (standard deviation,
skewness, kurtosis, third and first quartile difference, Stetson $K$ index (Stetson 1996) and von Neumann index (von Neumann 1941)) was used.
The list of parameters used in the second step is given in Tab.~2. Classification is again
performed with the Random Forest. {\it Recall} of this step, evaluated on the training set with 10-fold cross-validation is
89\% and {\it precision} of 88\%.

Taking into account the previously evaluated {\it recall} of the first (94\%) and second step (89\%),
one can estimate a theoretical {\it recall} value of the entire procedure to be 83\%. To verify this,
a sample of the OGLE-III binary stars from a different field (LMC532) is 
used as a test set. This set contains 962 objects, out of which none has been used in the training
set of the procedure, therefore they can be used to reliably evaluate the {\it recall} parameter of the classifier.
After performing a two step classification process, 761 objects have been correctly classified
as eclipsing binaries, resulting in a {\it recall} of 81.5\%, only slightly lower than the predicted one.

As a final verification, the light curves of all the eclipsing binary candidates returned by the automatic classifier are subjected to visual inspection.

\begin{table}[Hp]
\caption{Parameters obtained with the BLS algorithm used in the first and second step of the classification process.
For full description of the parameters see Kov{\'a}cs \etal (2002).}
\begin{tabular}{l l}
\\
\hline
\hline
Parameter & Description \\
\hline
\hline
$P$ &  Period \\
$SR$ & Signal Residue \\
$SN$ & Signal to Noise Ratio \\
$SDE$ & Signal Detection Efficiency \\
$D$ &  Depth of transit \\
$q$ &  Fraction of the phase in transit \\
$\chi^2$ & $\chi^2$ of the transit model fit \\
$RN$ & Red Noise \\
$WN$ &  White Noise \\
$SPN$ & Signal to Pink Noise Ratio \\
$P_{\rm{inv}}$ &  Period of inverse transit \\
$\chi^2_{\rm{inv}}$ & $\chi^2$ of the inverted transit model fit \\
$I$ & mean $I$-band magnitude \\
\hline
\end{tabular}
\end{table}

\begin{table}[Hp]
\caption{Statistical parameters used in the second step of the classification process only.}
\begin{tabular}{l l}
\\
\hline
\hline
Parameter & Description \\
\hline
\hline
$\sigma$ &  standard deviation \\
$\gamma_1$ & skewness \\
$\gamma_2$ & kurtosis \\
$Q_3 - Q_1$ & difference between third and first quartile \\
$\eta$ &  von Neumann index \\
$K$ &  Stetson $K$ index \\
\hline
\end{tabular}
\end{table}

\section{Eclipsing Binaries in the Magellanic System}

\begin{figure}[Hp]
\includegraphics[angle=270,width=62mm]{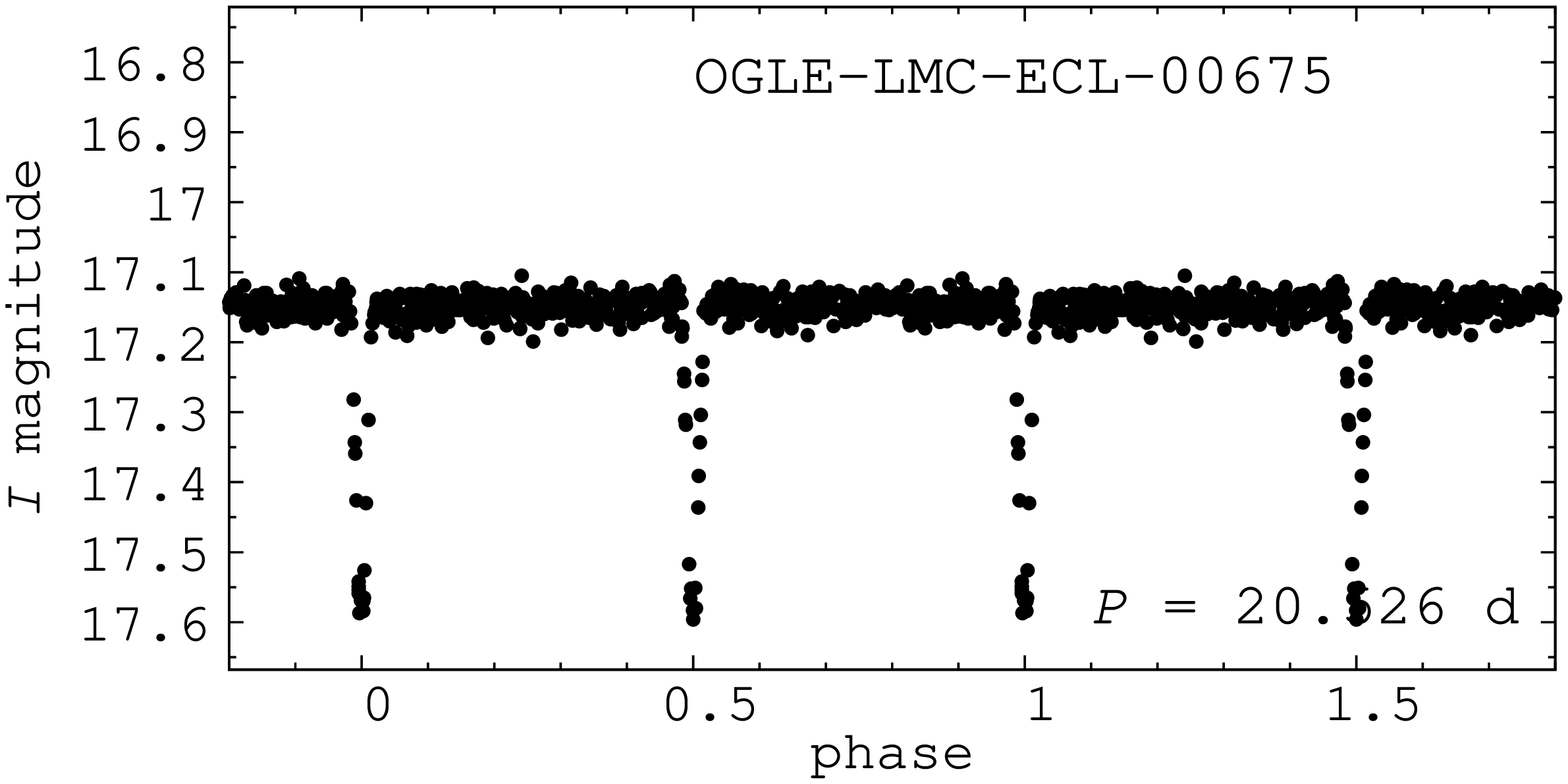}\hfill \includegraphics[angle=270,width=62mm]{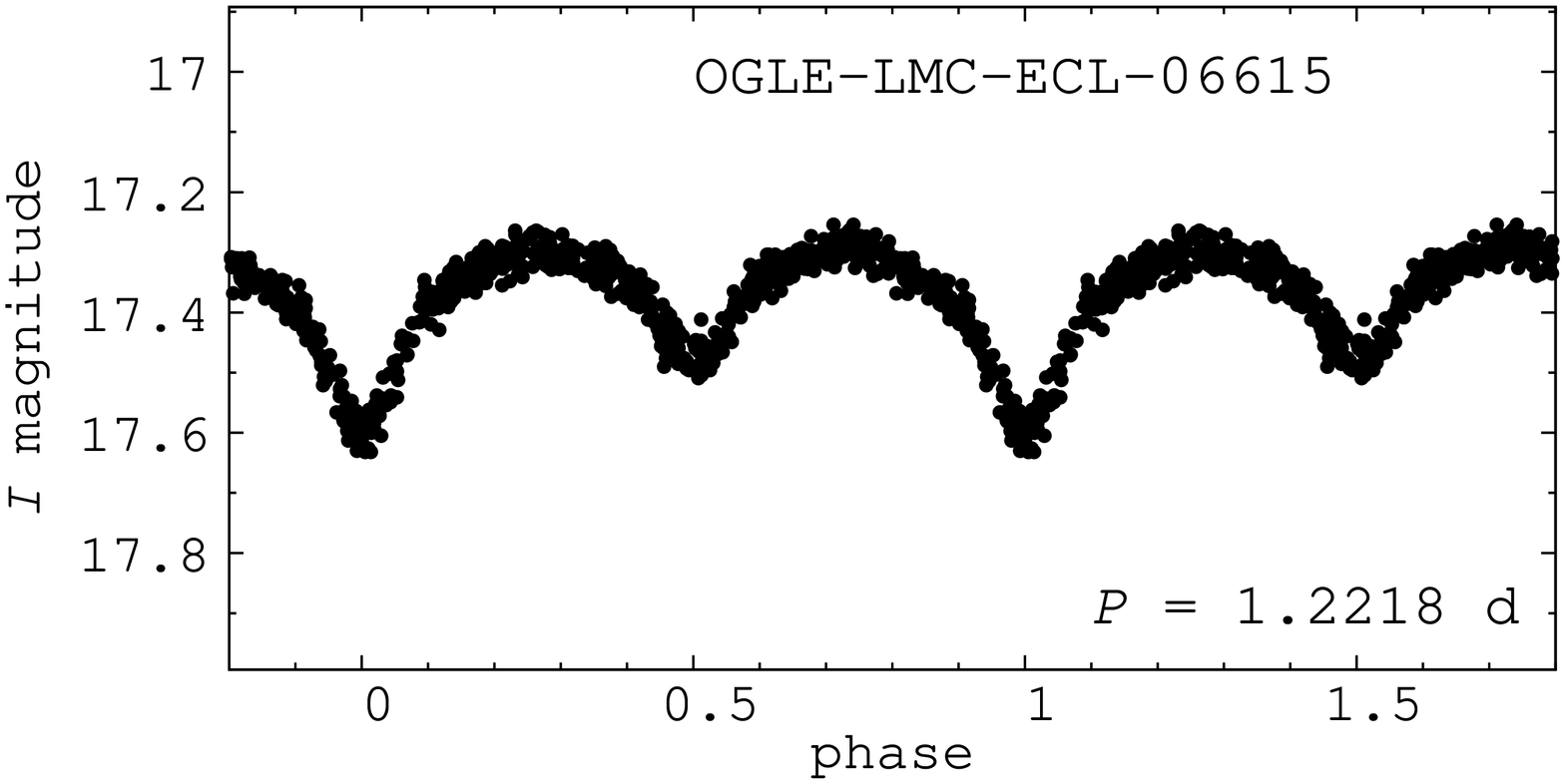} \\
\includegraphics[angle=270,width=62mm]{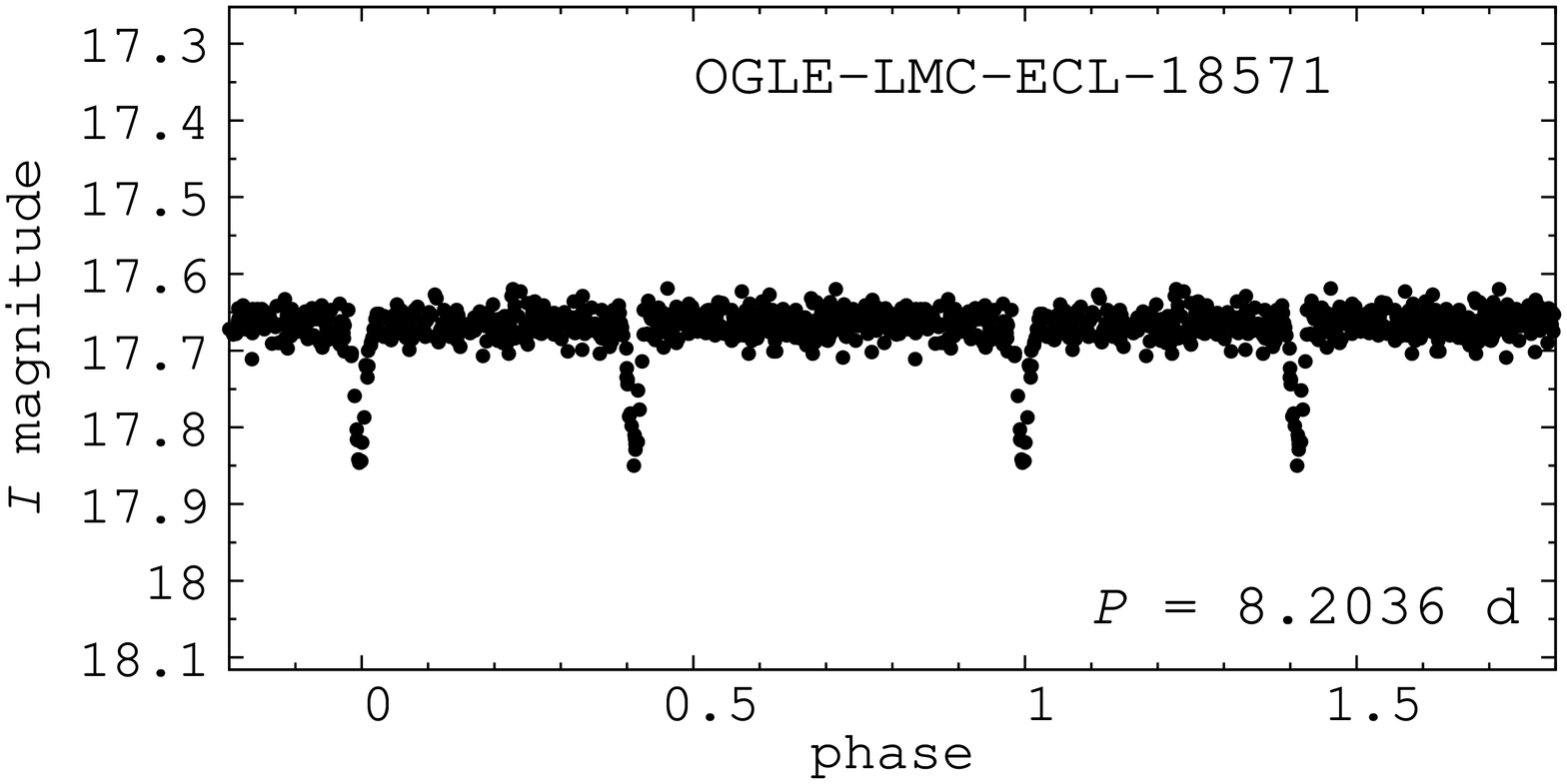}\hfill \includegraphics[angle=270,width=62mm]{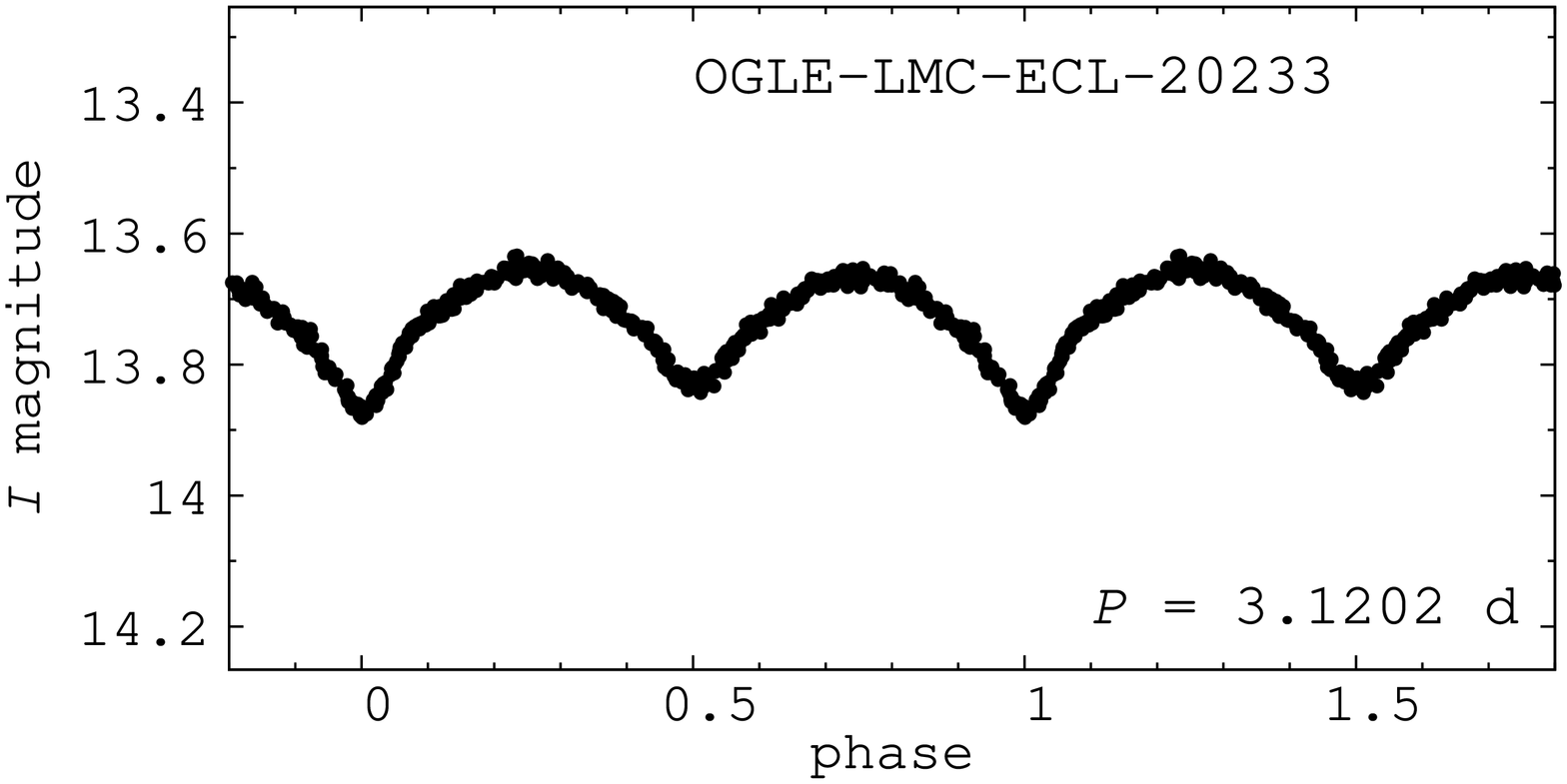} \\
\includegraphics[angle=270,width=62mm]{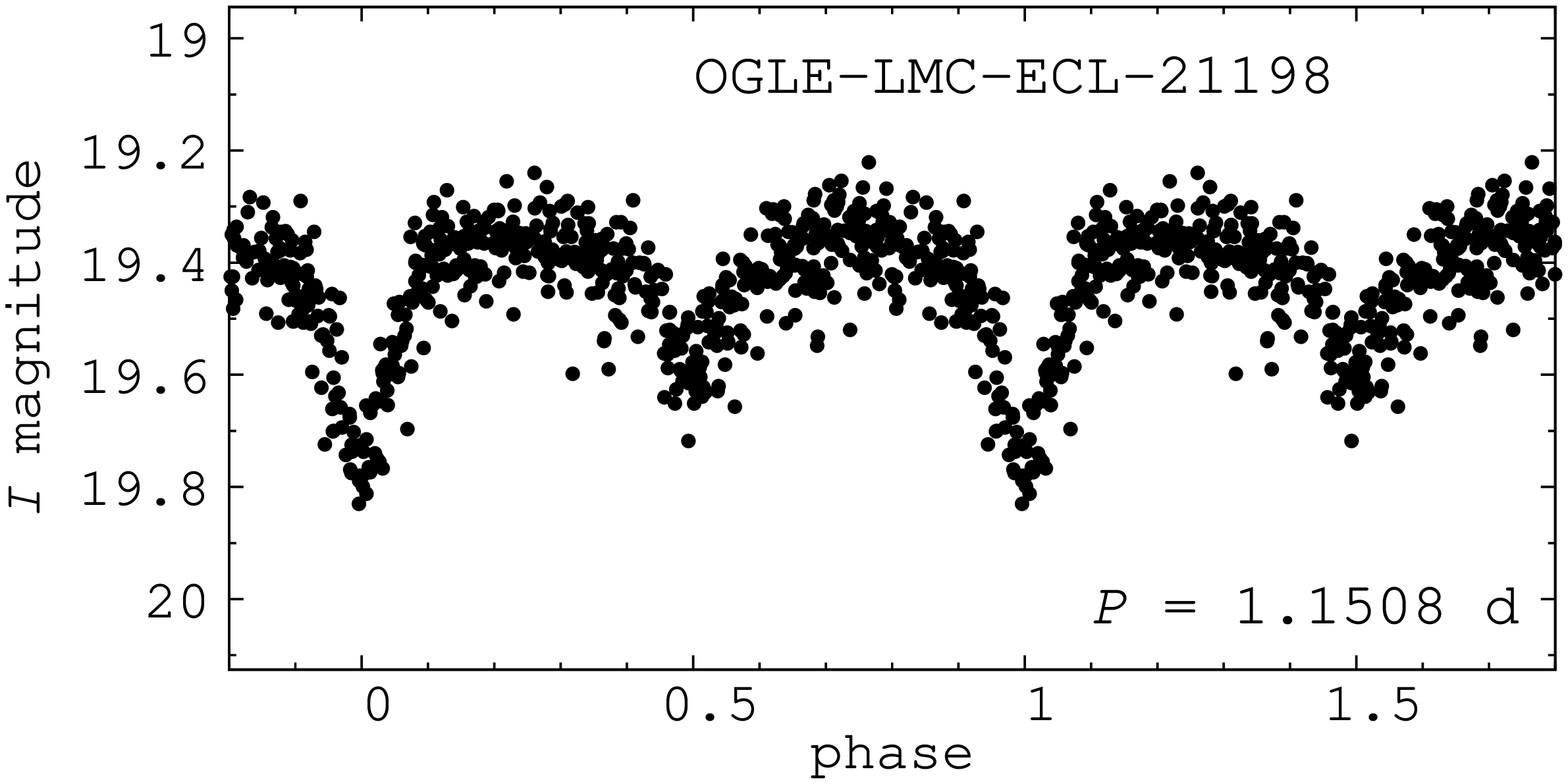}\hfill \includegraphics[angle=270,width=62mm]{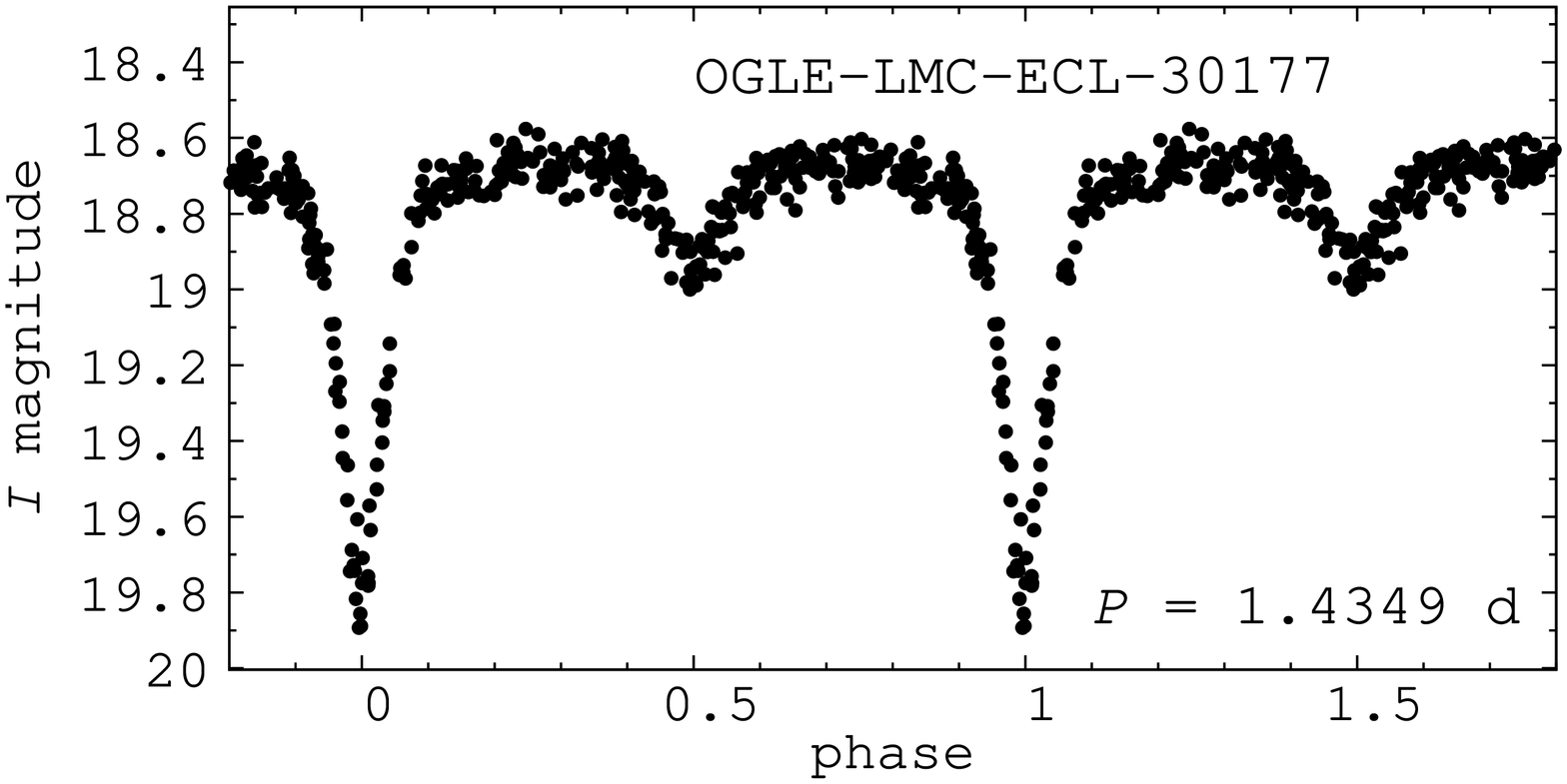} \\
\includegraphics[angle=270,width=62mm]{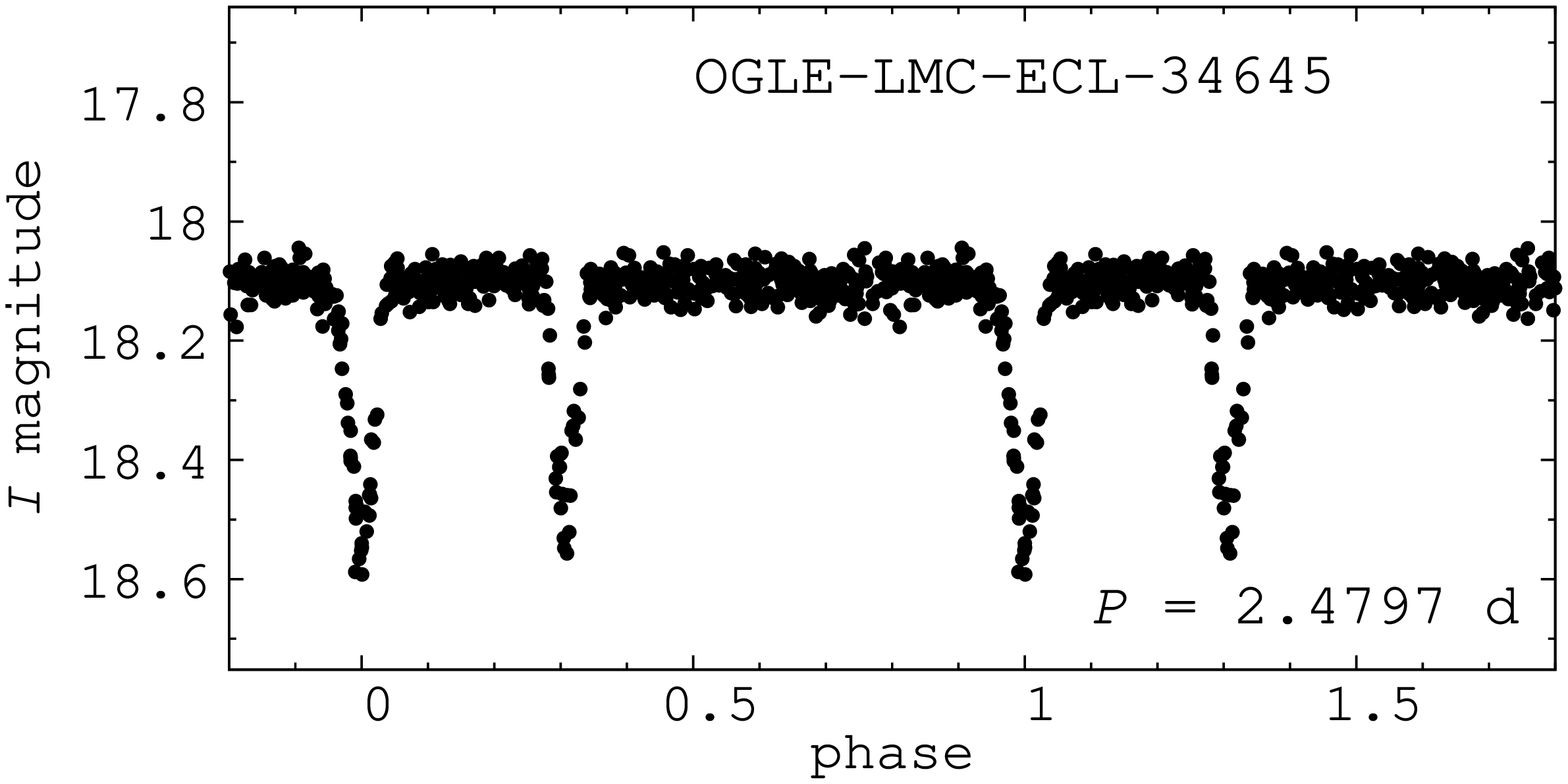}\hfill \includegraphics[angle=270,width=62mm]{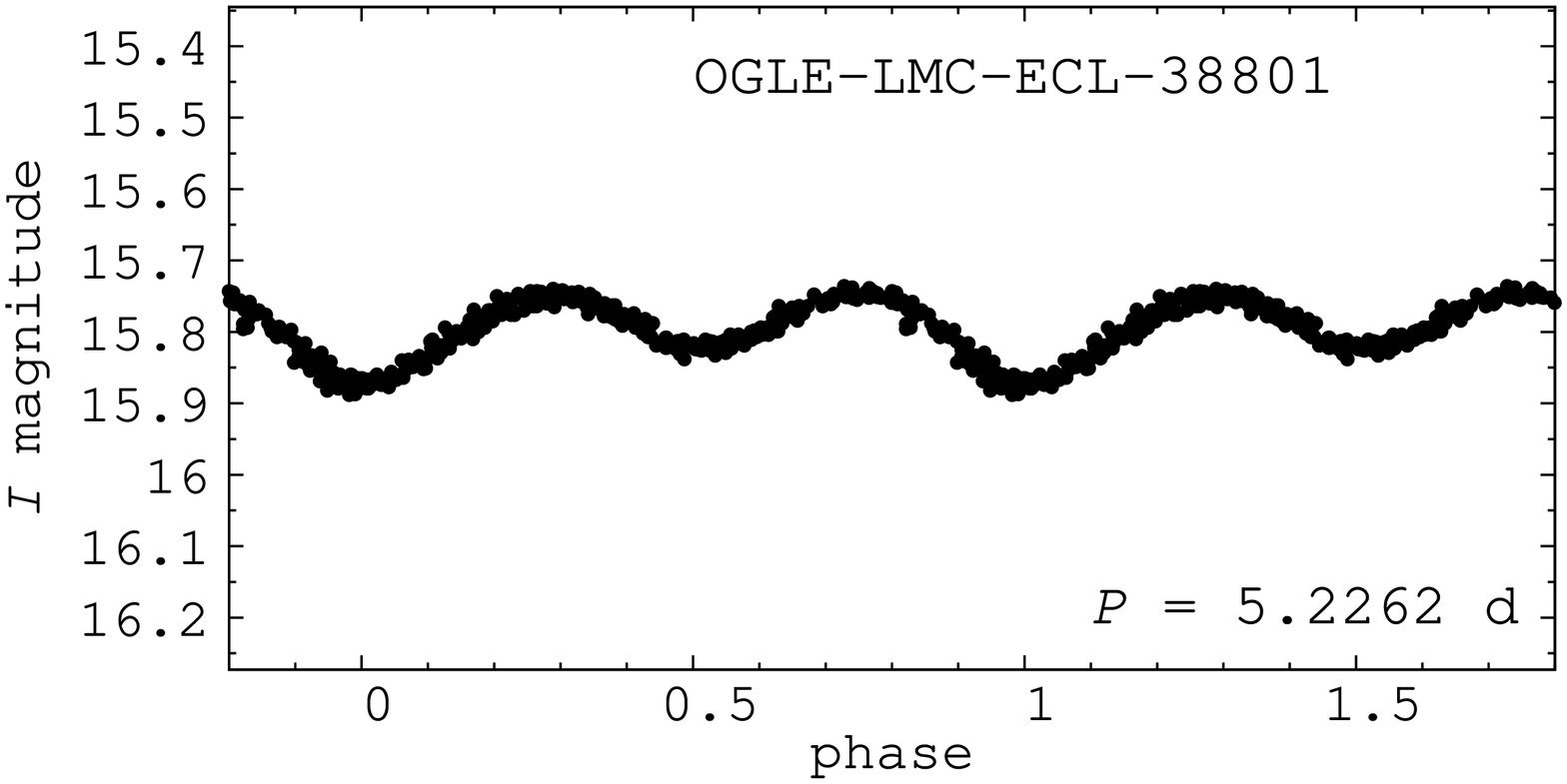} \\
\includegraphics[angle=270,width=62mm]{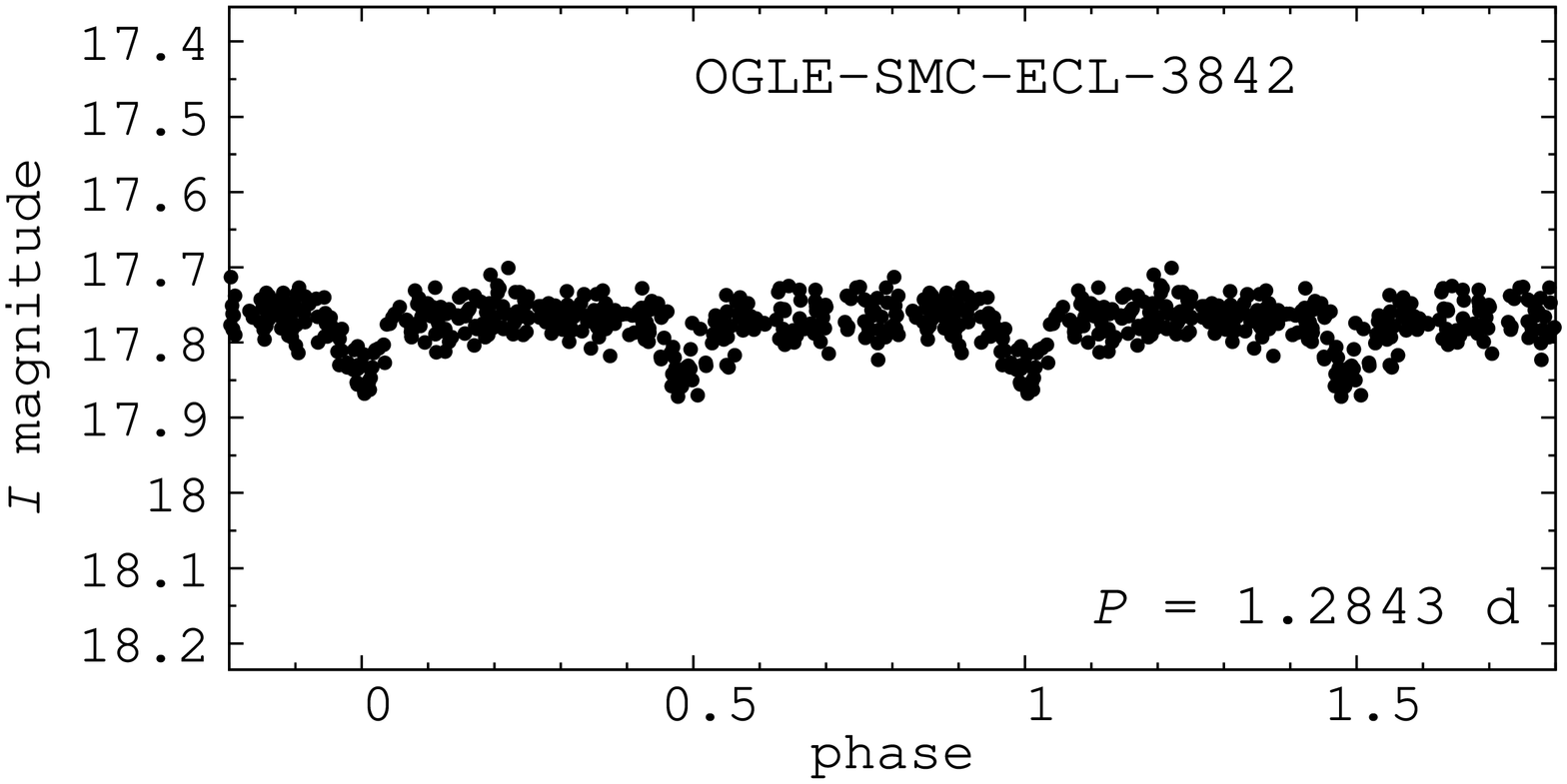}\hfill \includegraphics[angle=270,width=62mm]{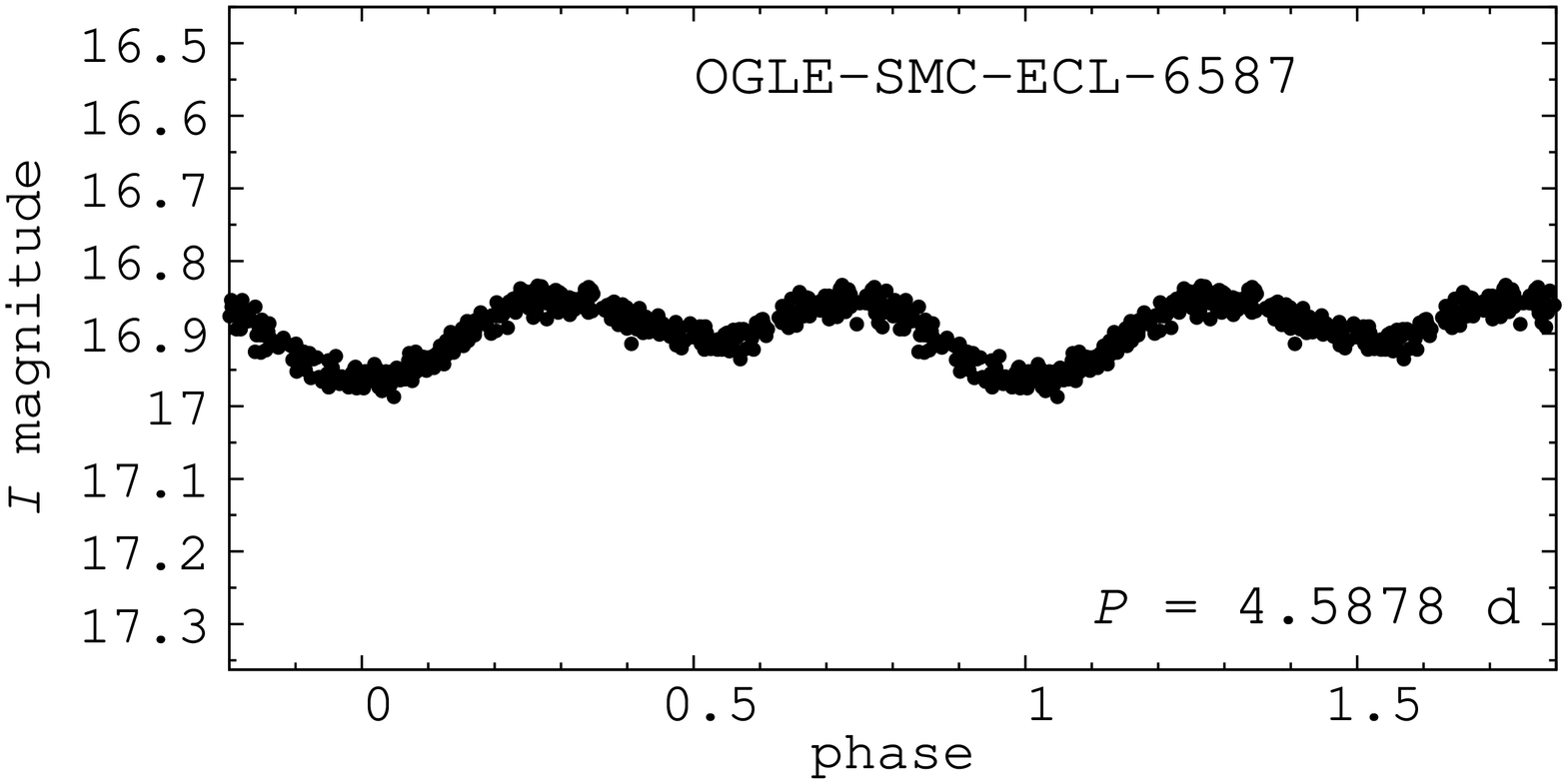} 
\FigCap{Examples of phased light curves of eclipsing binary systems in the Magellanic System.}
\end{figure}

The collection of the eclipsing binaries in the Magellanic System presented in this work consists of
objects from the OGLE-III catalogs (Poleski \etal 2010,  Graczyk \etal 2011, Pawlak \etal 2013), Classical 
and Type II Cepheids in the eclipsing systems (Udalski \etal 2015b, Soszy{\'n}ski \etal 2008, 2010),
as well as $16\;374$ newly discovered objects making a total of $48\;605$. Out of the entire sample, $40\;204$ binary systems are classified as belonging
to the LMC and $8\;401$ to the SMC. The division between the LMC and SMC is set on the celestial meridian of $2.8^h$, the same
as the one adopted by Soszy{\'n}ski \etal (2016) for RR Lyrae stars in the Magellanic System.

Eclipsing binaries are traditionally divided into three subtypes: contact, semi-detached, and detached. 
The identification of likely contact systems can by done relatively easily based on the morphology of the light curve.
These objects have a smooth transition from the eclipse to the out-of-eclipse phase and two minima of the same or very similar
depth, due to the fact that the systems are in thermal equilibrium and both components have the same surface temperature.
The situation becomes more difficult for semi-detached systems, which are often hard to distinguish from the close but
still detached ones, using the photometric data only. Therefore, in this work we divide our sample of eclipsing binaries
into contact and non-contact, with the latter ones containing both semi-detached and detached objects. We also extract
a sample of non-eclipsing, ellipsoidal binaries.

Example light curves of the objects from the catalog are presented in Fig.~1.
The entire collection is available to the astronomical community in the OGLE Internet Archive, accessible via
FTP sites or a web interface: 

\begin{center}
{\it ftp://ftp.astrouw.edu.pl/ogle/ogle4/OCVS/lmc/ecl/}\\
{\it ftp://ftp.astrouw.edu.pl/ogle/ogle4/OCVS/smc/ecl/}\\
{\it http://ogle.astrouw.edu.pl}
\end{center}

The collection contains the list of parameters of the objects (coordinates, periods, {\it I}- and {\it V}-band out-of-eclipse magnitudes, epoch of primary eclipse, 
depth of primary and secondary eclipse), 
as well as {\it I}- and {\it V}-band time series photometry collected during the fourth phase of the OGLE survey. This photometry can be combined with the OGLE-III data 
(Graczyk \etal 2011, Pawlak \etal 2013), though in individual cases it may be necessary to adjust magnitudes and amplitudes as they may
vary slightly due to blending and crowding. Cross-identification with the General Catalog of Variable Stars (Artyukhina \etal 1995) is also provided.

\section{Discussion}

The {\it I}-band magnitude range of objects presented in our sample is from 13 to 20.4~mag. The histogram
of the magnitudes is presented in Fig.~2. The distribution has a maximum at 19~mag. The completeness of
the search drops rapidly for systems fainter than 19.6~mag, as the noise of photometric measurements 
increases, making it difficult to detect eclipsing stars.

The distribution of the orbital periods is presented in Fig.~3. The majority of systems have period within the range of 1-10~d.
This is different from the typical distribution for Galactic binary stars (\eg Pietrukowicz \etal 2013), where the
maximum falls to a period shorter than 1~d. This is likely an observational bias. Large fraction of short-period binaries consists of W~UMa type stars, which are 
in most cases fainter than 21~mag and they cannot be detected in the OGLE data. The absence of these objects in our sample
shifts the maximum of the histogram toward the longer periods.

\begin{figure}[htb]
\centering{
\includegraphics[width=58mm, angle=270]{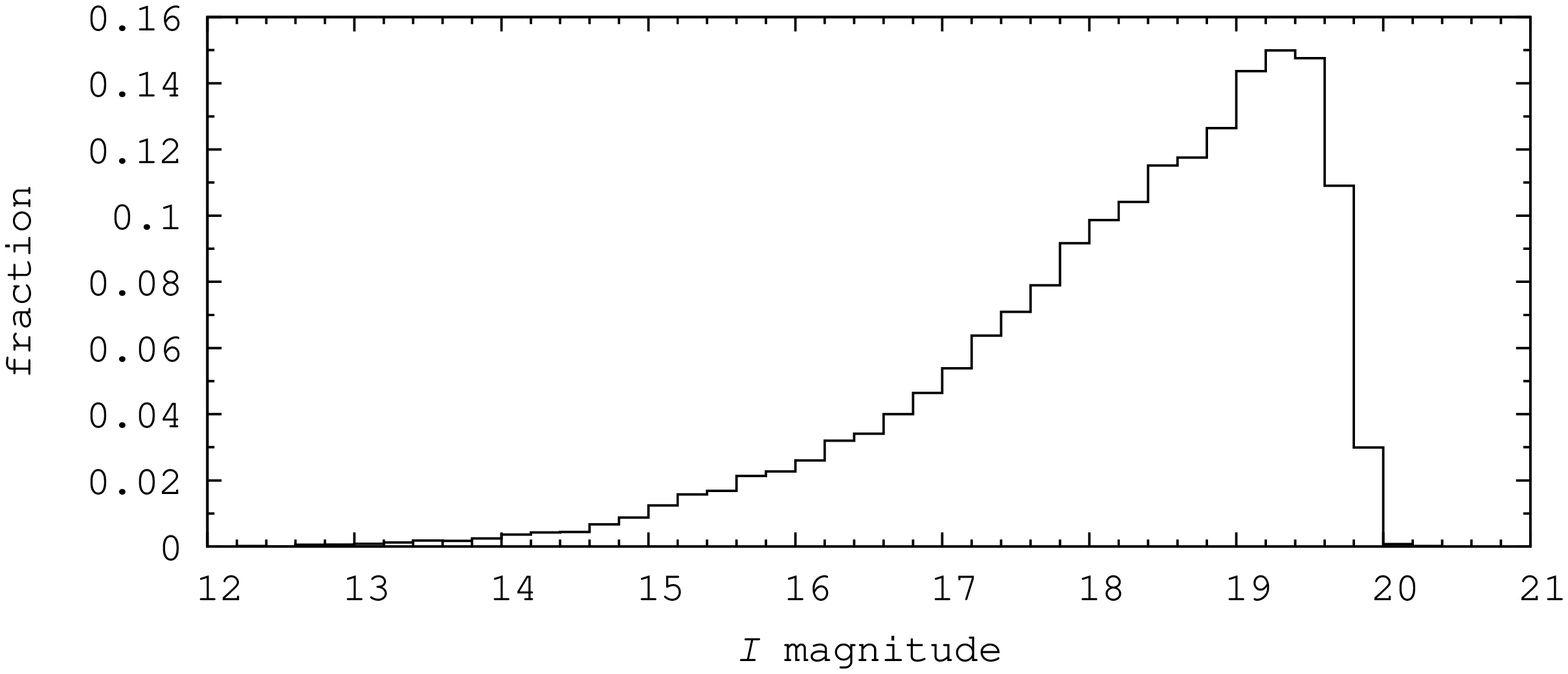}}
\FigCap{Histogram of the out-of-eclipse magnitudes of binary systems in the Magellanic System.}
\end{figure}

\begin{figure}[htb]
\centering{
\includegraphics[width=58mm, angle=270]{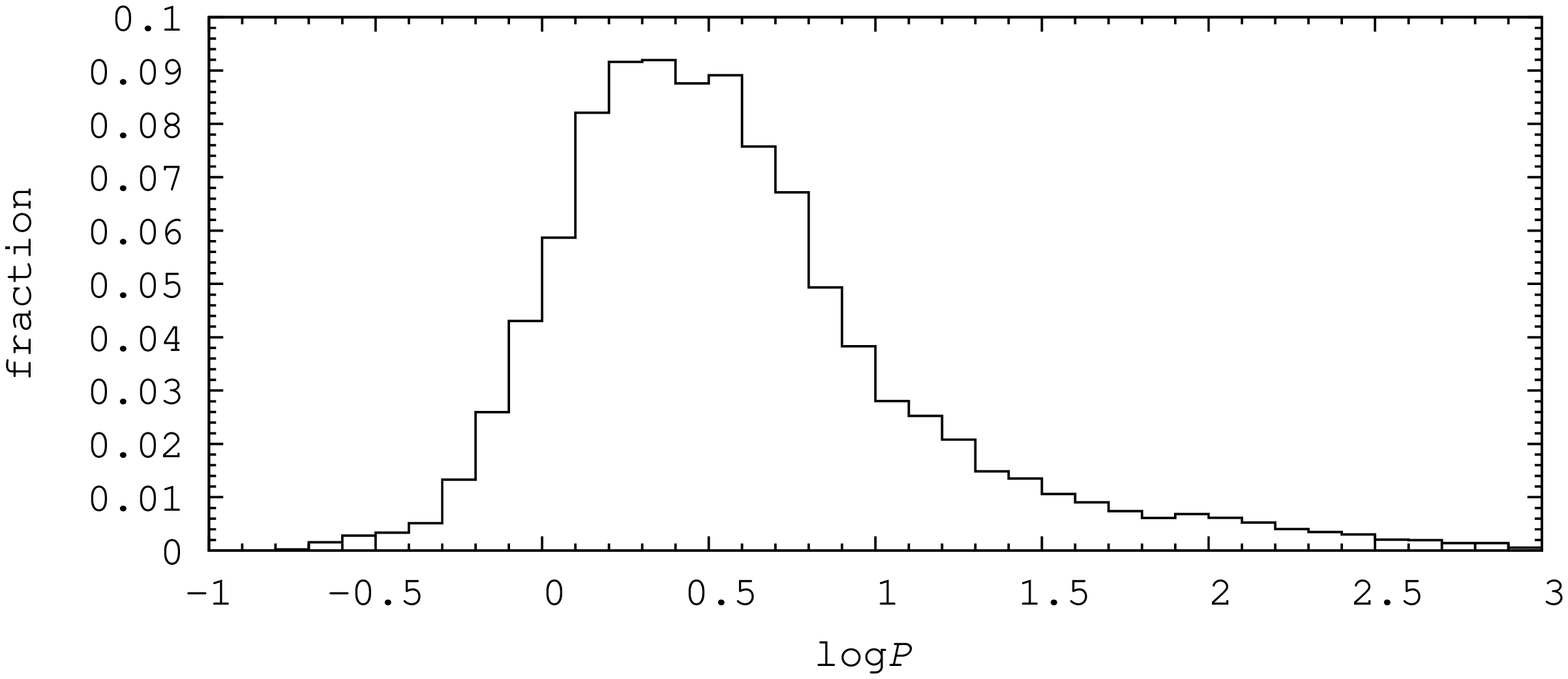}}
\FigCap{Histogram of the orbital periods of binary systems in the Magellanic System.}
\end{figure}

The large area of the sky covered by OGLE-IV allows us to study the distribution of the eclipsing systems in the whole Magellanic Clouds.
Distribution of objects from the studied sample on the sky is presented in Fig.~4. 
While identified eclipsing binaries are likely to be a mixture of populations, they seem to trace the bar and spiral arms of the LMC and group
around the center of SMC, which resembles more the distribution of the young population (Soszy{\'n}ski \etal 2015) than the old one (Soszy{\'n}ski \etal 2016). 
It is also consistent with the young population maps for the Magellanic Bridge region (Skowron \etal 2014).

Location of the identified systems in the color-magnitude diagram (Fig.~5 and Fig.~6) shows that most of them are main sequence stars of A and B type,
thus they belong to the young population. However, there is also a significant sample of red giant binaries, especially among the ellipsoidal systems.

\begin{figure}[Htb]
\centering{
\includegraphics[width=127mm, angle=0]{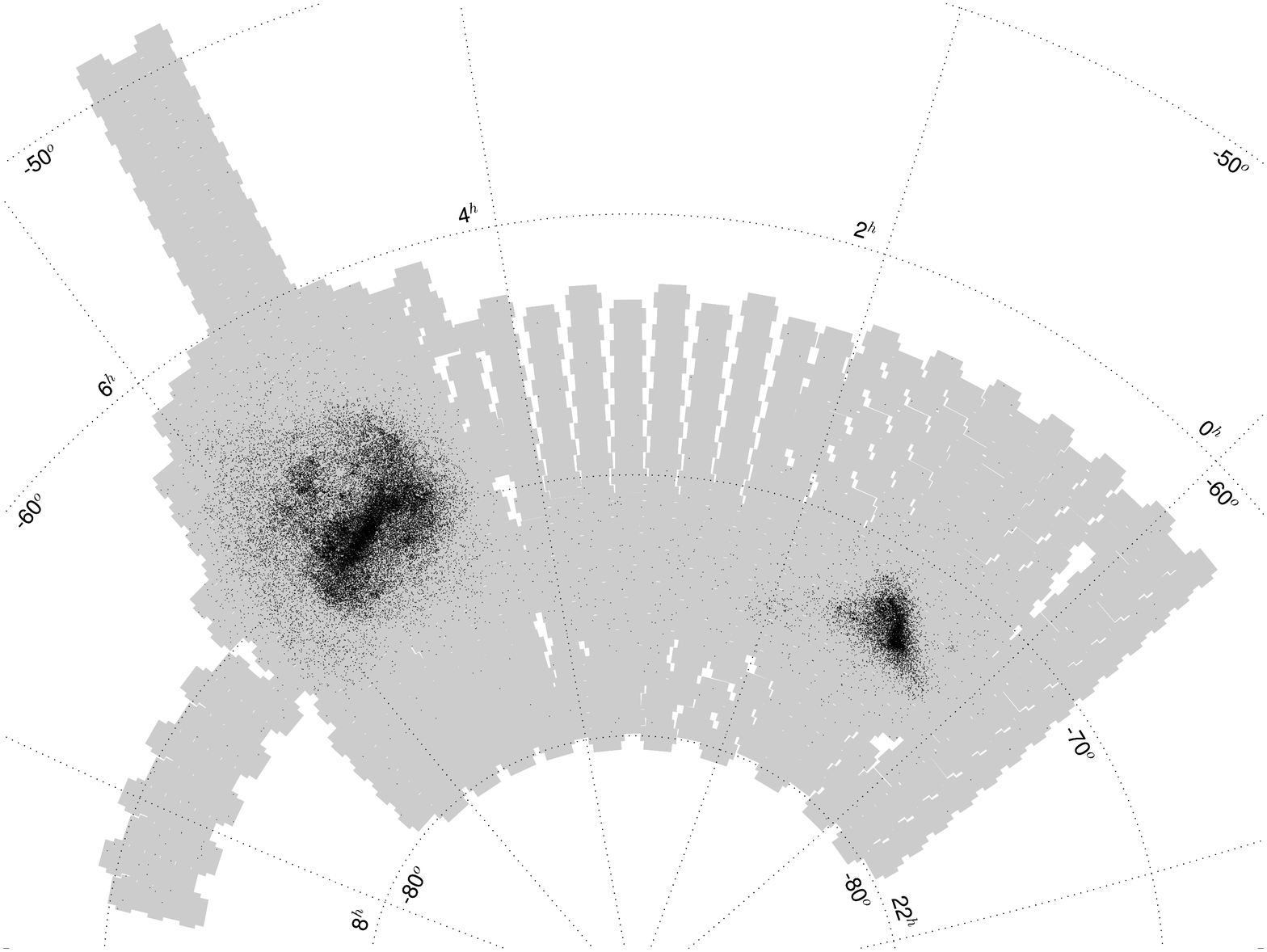}}
\FigCap{The spatial distribution of binary stars in the area of the Magellanic System. Gray area shows the sky coverage of the OGLE-IV fields.}
\end{figure}

\begin{figure}[Htb]
\centering{
\includegraphics[width=92mm, angle=270]{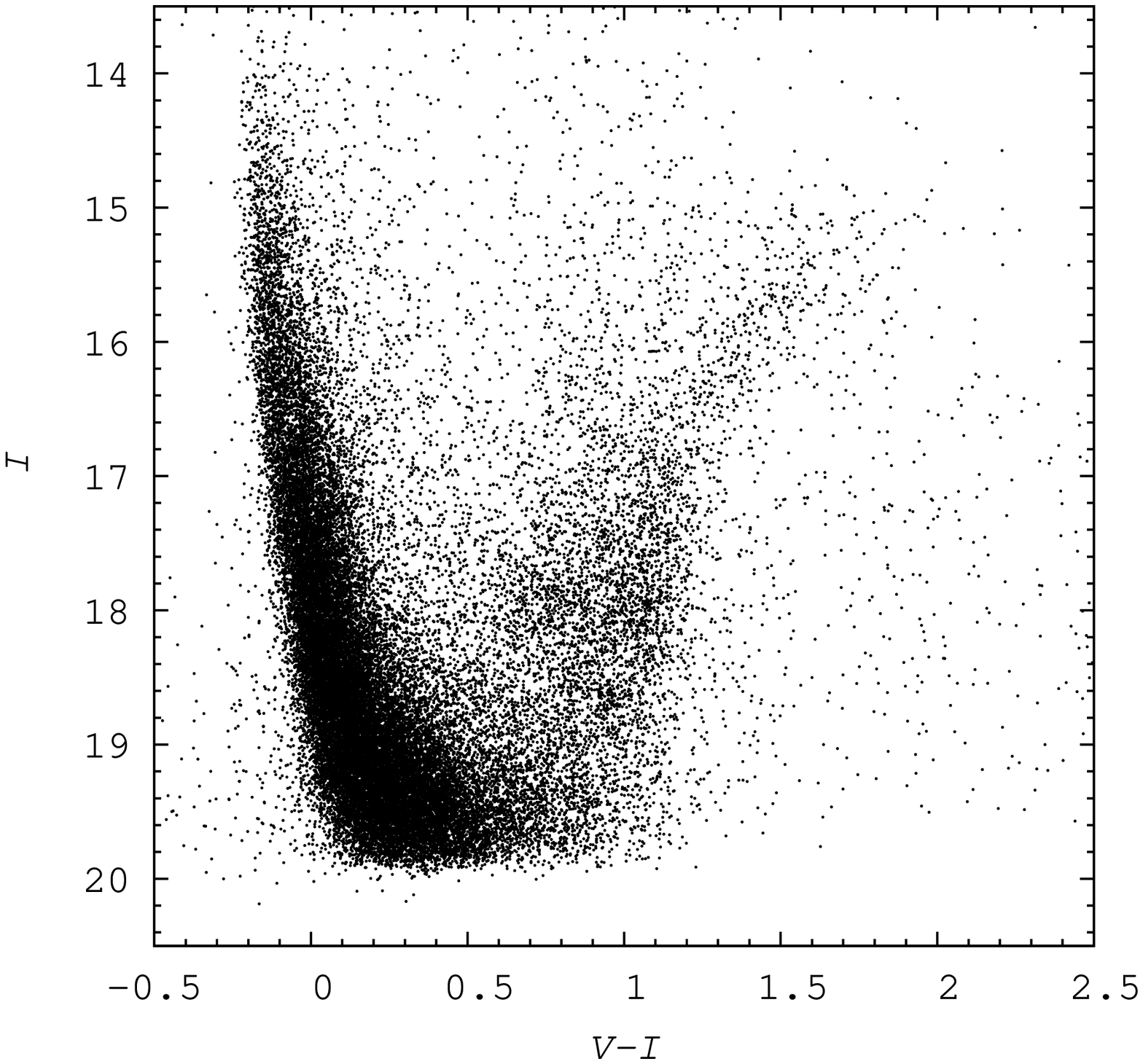}}
\FigCap{Color-magnitude diagram for the binary systems in the LMC.}
\end{figure}

\begin{figure}[Htb]
\centering{
\includegraphics[width=92mm, angle=270]{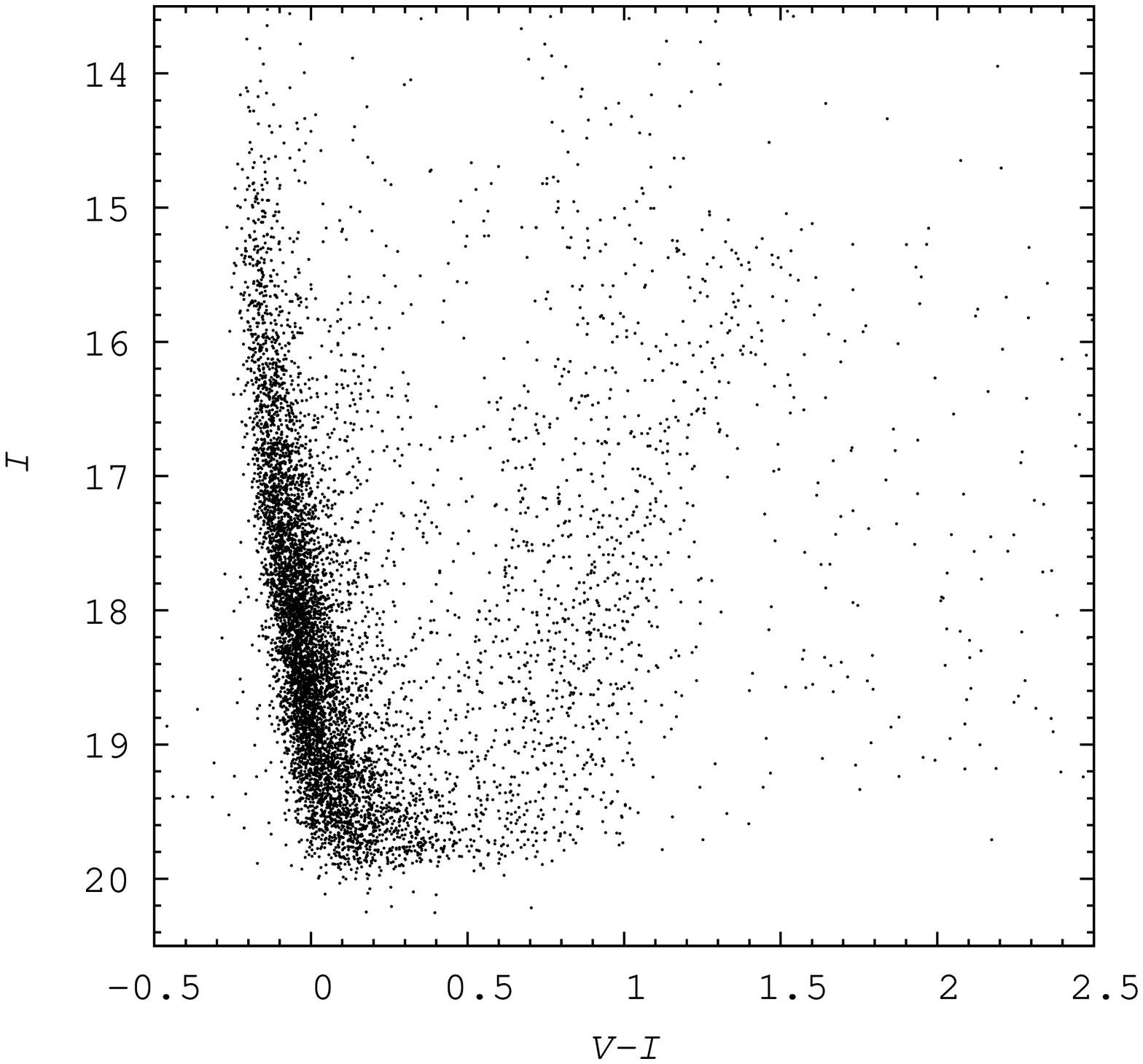}}
\FigCap{Color-magnitude diagram for the binary systems in the SMC.}
\end{figure}

\section{Eclipsing Binary with a Type II Cepheid}

OGLE-LMC-ECL-26653 (Fig.~7) is an eclipsing system with one of the components being a type~II Cepheid. This object was not included in 
the OGLE-III collection of type~II Cepheids in the LMC (Soszy{\'n}ski \etal 2008), since it is located outside the OGLE-III area.
The orbital period of the system is $P_{\rm orb} = 242.4338$~d and the pulsation period of the Cepheid
is $P_{\rm puls} = 9.39258$~d. The shape of the pulsation light curve shows
that this object belongs to the so called peculiar W~Virginis stars -- a subtype
of type~II Cepheids distinguished for the first time by Soszy{\'n}ski \etal (2008).
The automatic search procedure identified this object as an eclipsing binary candidate, showing the capability of the 
method to detect even such peculiar objects. OGLE-LMC-ECL-26653 is the fourteenth type~II Cepheid
in an eclipsing binary system discovered in the Magellanic System by the OGLE survey. 

\begin{figure}[Htb]
\includegraphics[angle=270,width=62mm]{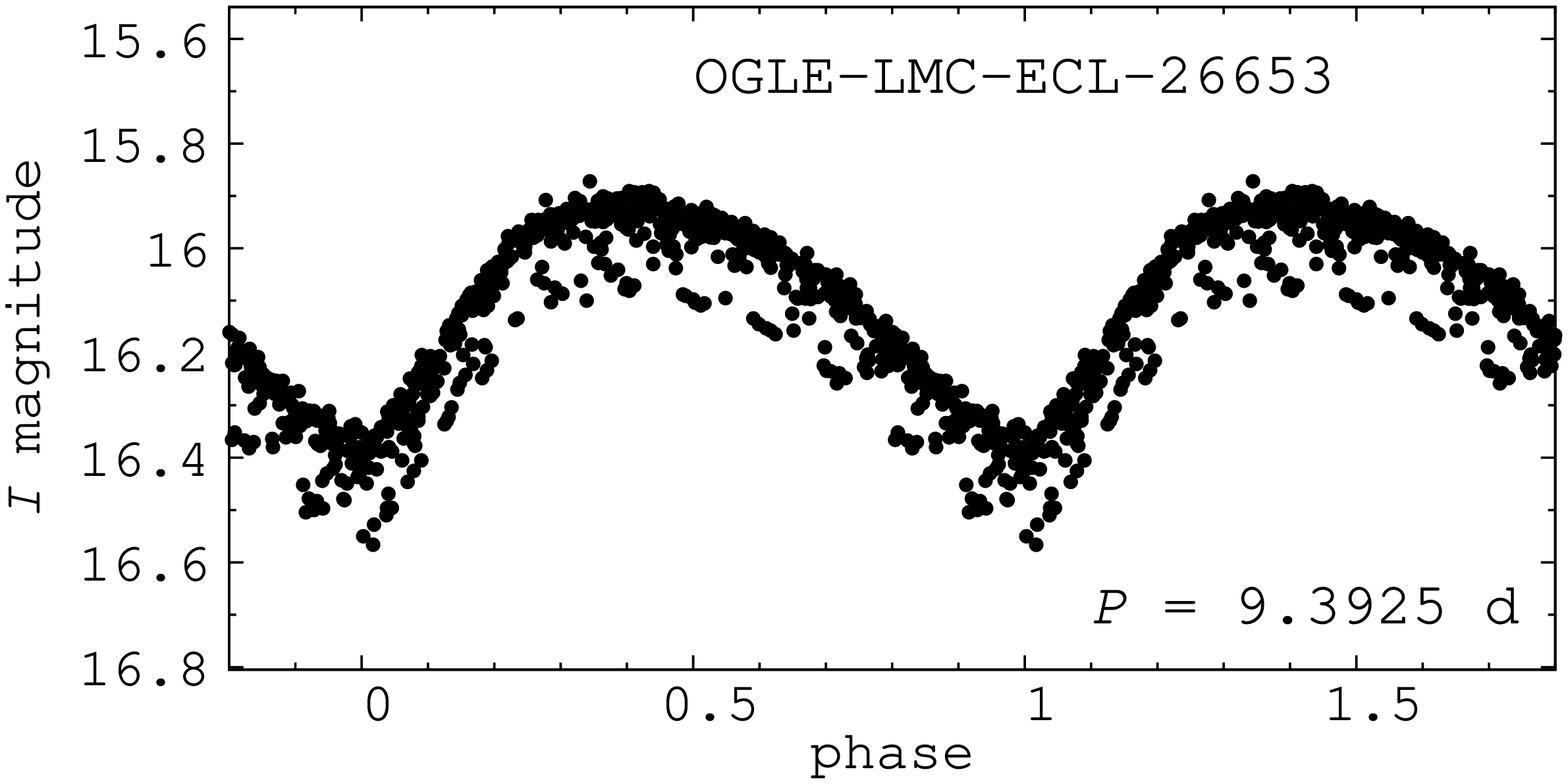}\hfill \includegraphics[angle=270,width=62mm]{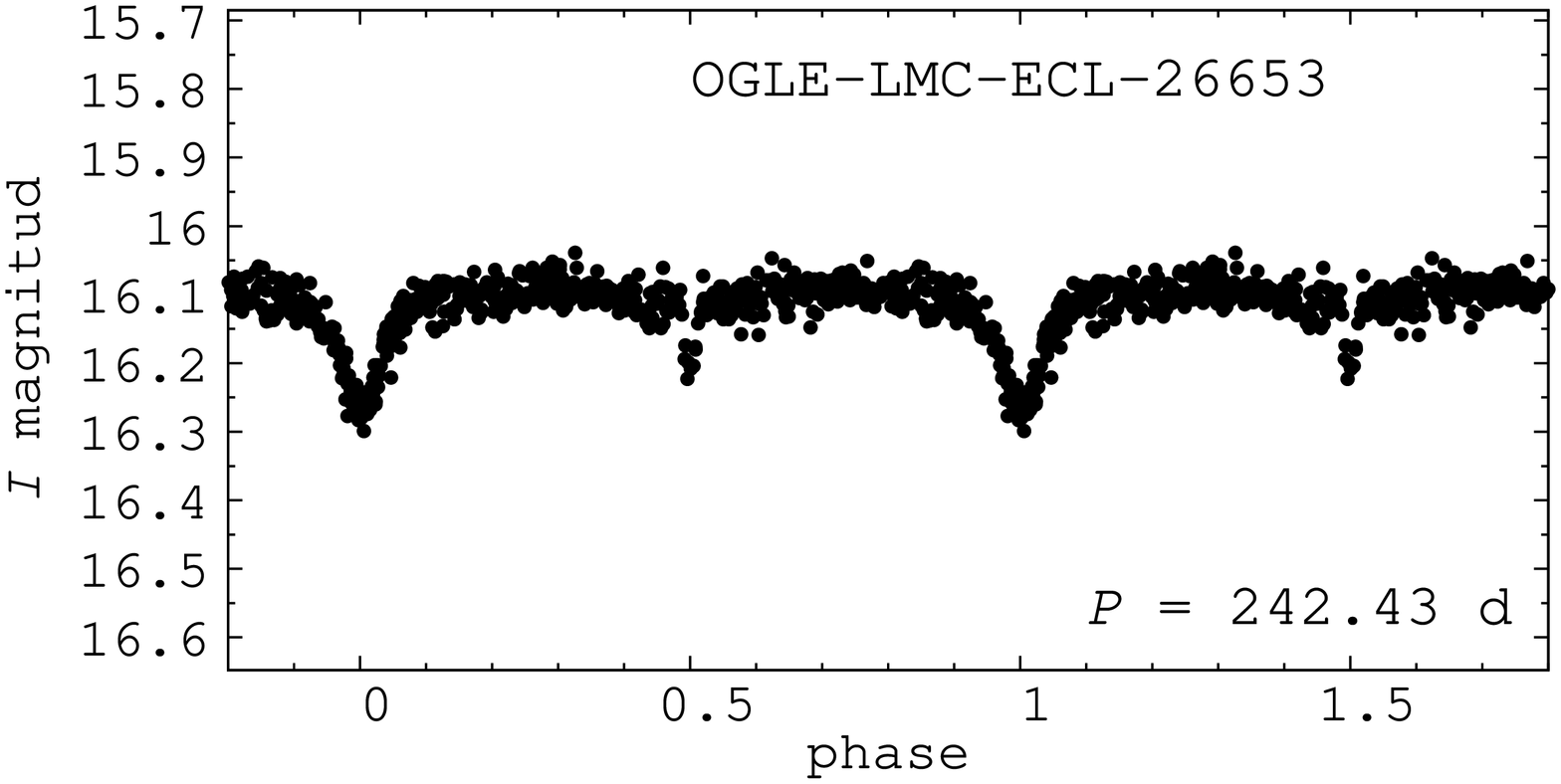}
\FigCap{OGLE-LMC-ECL-26653 light curve phased with the pulsating period (left panel) and with the orbital period (right panel, pulsating variability prewhitened).}
\end{figure}

\section{Summary}

The OGLE-IV collection of binary systems in the Magellanic System presented in this work contains $48\;605$ objects, 
which makes it the largest catalog of binary stars in this stellar environment. $40\;204$ objects
were identified as belonging to the LMC and $8\;401$ to the SMC. We provide a series of parameters
as well as {\it I}- and {\it V}-band time-series photmetry of the identified objects. 

The search was performed using a two-step machine learning method based on the Random Forest algorithm and the final
verification of each of the candidates was done with visual inspection.
The overall completeness of the search is estimated to be over 80\%.

The catalog reaches up to 20.4~mag in
$I$-band, with the relatively high completeness up to 19.6~mag. The spatial distribution of objects from the sample
as well as their position in the color-magnitude diagram suggest that most of them are main sequence stars belonging to the young population.

\Acknow{We would like to thank Profs. M. Kubiak and G. Pietrzy{\'n}ski, former members of the OGLE team, for their contribution
to the collection of the OGLE photometric data over the past years. 

This work has been supported by Polish National Science Center grants PRELUDIUM no. 2014/13/N/ST9/00075 and OPUS no. 2011/03/B/ST9/02573. 
MP is supported by Polish National Science Center under the ETIUDA grant no. 2016/20/\\T/ST9/00170. 
The OGLE project has received funding from the Polish National Science Center grant MAESTRO no. 2014/14/A/ST9/00121 to AU.}

\end{document}